\documentclass[10pt,twocolumn,letterpaper]{article} 

\usepackage{avss}
\usepackage{times}
\usepackage{epsfig}
\usepackage{url}
\usepackage{hyperref}
\usepackage{graphicx}
\usepackage{amssymb}

\usepackage{subfig}
\usepackage{amsthm,amsmath}
\usepackage{mathrsfs}
\usepackage{multirow}
\usepackage{makecell}
\usepackage{xcolor}
\usepackage{pifont}
\usepackage{tabularx}
\usepackage{booktabs}
\usepackage{adjustbox}
\usepackage{enumitem}
\usepackage{pifont}
\avssfinalcopy 

\def\avssPaperID{128} 

\ifavssfinal\fi
\begin{document}

\title{Data-driven RF Tomography via Cross-modal Sensing and Continual Learning}

\author{Yang Zhao, Tao Wang, Said Elhadi\\
Harbin Institute of Technology, Shenzhen\\
University Town of Shenzhen, Shenzhen, Guangdong Province, China\\
}

\maketitle

\begin{abstract}
Data-driven radio frequency (RF) tomography has demonstrated significant potential for underground target detection, due to the penetrative nature of RF signals through soil.
However, it is still challenging to achieve accurate and robust performance in dynamic environments. In this work, we propose a data-driven radio frequency tomography
(DRIFT) framework with the following key components to reconstruct cross section images of underground root tubers, even with significant changes in RF signals. First, we design a cross-modal sensing system with RF and visual sensors, and propose to train an RF tomography deep neural network (DNN) model following the cross-modal learning approach. Then we propose to apply continual learning to automatically update the DNN model, once environment changes are detected in a dynamic environment. Experimental results show that our approach achieves an average equivalent diameter error of 2.29 cm, 23.2\% improvement upon the state-of-the-art approach. Our DRIFT code and dataset are publicly available on https://github.com/Data-driven-RTI/DRIFT.
\end{abstract}
\vspace{-1.5em}


\section{Introduction}
As the development of deep learning and embedded sensing techniques, various sensors and DNN models have been proposed to detect and monitor crop above-ground phenotypic traits, e.g., leaf area index, in crop yield prediction, smart breeding, and other smart agriculture applications~\cite{roitsch2019new, gano2024drone}. 
However, underground root sensing remains an important problem largely under-researched, especially for root vegetables and crops bearing starchy tuberous roots, e.g., potato.
Ground penetrating radar~(GPR) can detect underground targets using RF signals. However, a radar sensor has a distance loss inversely proportional to distance to the fourth power ($1/D^4$), and a higher frequency band generally leads to lower penetration capability. 
Recent studies show that data-driven methods have achieved state-of-the-art~(SOTA) performance in networked RF sensing. For example, a data-driven RF tomography approach is proposed to use commercial-off-the-shelf wireless nodes and build upon the radio tomographic imaging framework~\cite{wilson2010see, zhao2014robust} to detect underground potato tubers\cite{wang2024demo}. However, data-driven RF tomography for detecting underground targets has the following issues. 

\begin{figure}[!t]
    \centering
    \includegraphics[width=\linewidth]{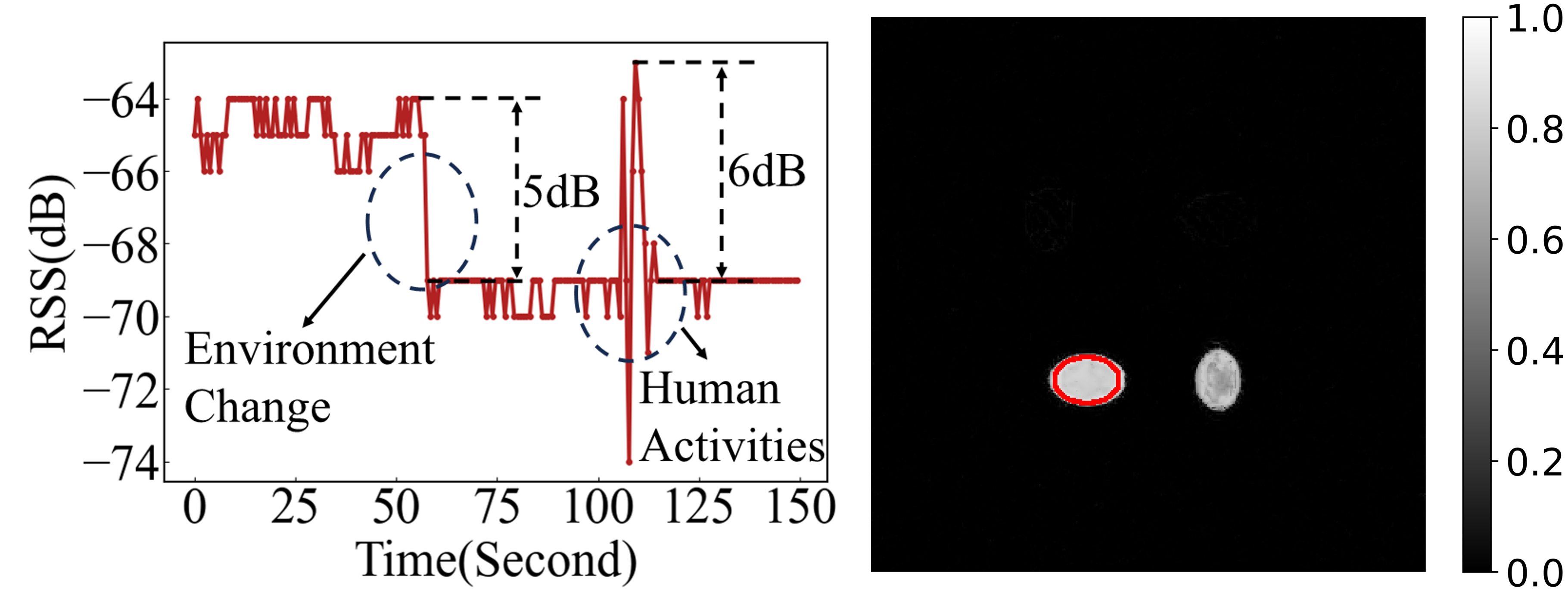}
    \caption{Changes in RF signals and their effects on an RF tomography detection system: RSS changes due to human activities and environmental layout changes (left), Degradation in RF tomography (red circle indicates the cross section ground-truth of a potato tuber, right).} 
    \label{fig:envs}
\end{figure}

First, environmental changes can cause significant performance degradation of an RF tomography detection system due to the multi-path effects. As shown in the left panel of Fig.~\ref{fig:envs}, human activities can produce significant fluctuations in received signal strength (RSS) measurements of an RF tomography system~\cite{zhao2014robust}. An environmental layout change, e.g., door opening, can also lead to RSS variations up to 5 dB due to changes in the multi-path effects~\cite{fu2020environment}. As a result, the false positive blob next to the tuber ground-truth in the right panel of Fig.~\ref{fig:envs} indicates the “fingerprint” left from the previous environmental layout. 
Second, DNN-model based RF tomography has demonstrated significant potential for underground root tuber sensing, given sufficient well-annotated data at a particular environment. However, it is challenging to annotate large amounts of data, due to the non-line-of-sight issue for underground object detection. Although GPRs and RF sensor networks have the capability of penetrating soils, it is time-consuming and labor-intensive to bury and dig the objects underground, e.g., potato tubers, and measure their dimensions and other properties manually. 

\begin{figure*}
    \centering
    \includegraphics[width=0.8\linewidth]{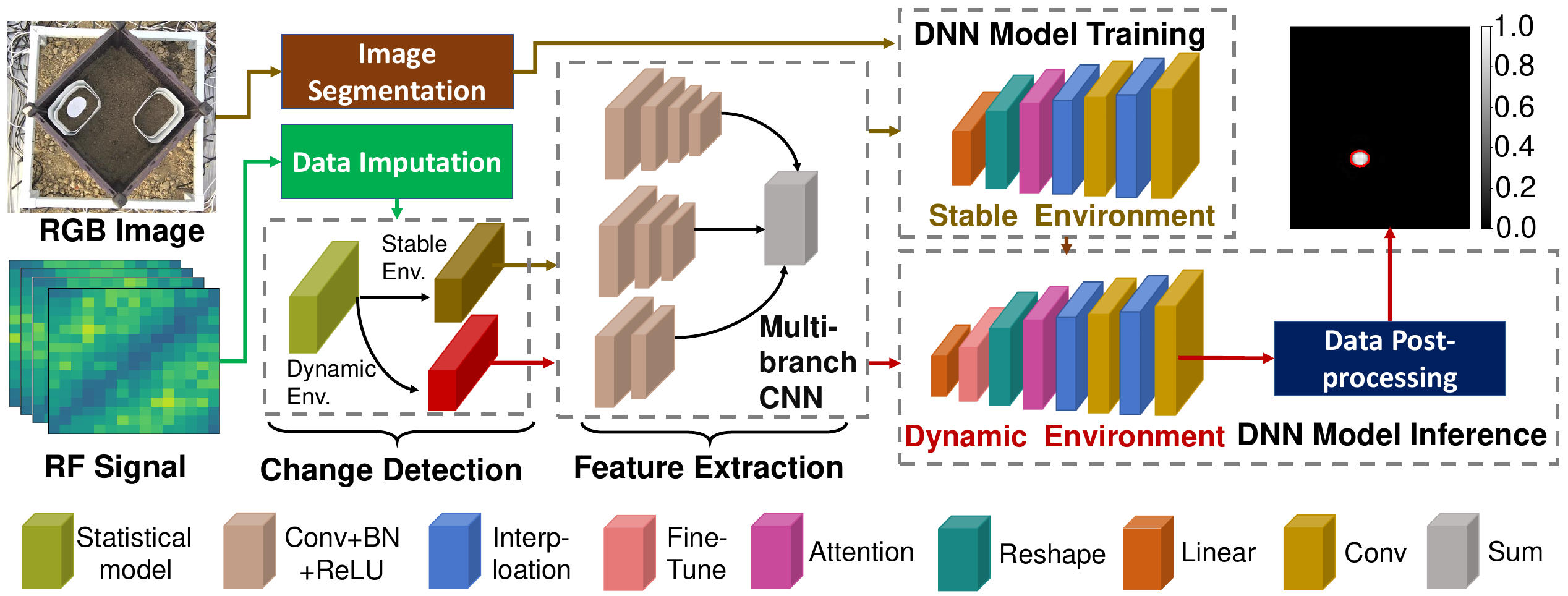}
    \caption{Overview of the DRIFT framework with change detection, multi-branch CNN, fine-tuning, and other modules.}
    \label{fig:architecture}
\end{figure*}

To tackle these challenges, we propose a statistical model-based environmental change detector and a continual learning framework that integrates a one-shot fine-tuning strategy with an encoder-decoder network to achieve robust imaging in dynamic environments. Specifically, the variance of RSS measurements within a window is modeled and used as the indicator of environmental changes. 
If no changes in the environment are detected, the recorded RF data will be combined with the segmentation results from our visual system to train our RF tomography DNN model. Our DNN model includes a multi-branch CNN module, which is used as the encoder to extract high-dimensional features from RSS measurements. Since the attention mechanism can effectively extract noise information from complex background, our DNN model also includes a novel decoder with attention and convolutional layers called ANC to filter out noise unrelated to root tubers. If changes are detected by our statistical model, we use the one-shot fine-tuning method~\cite{yang2023one} on the most recent RSS measurements to fine-tune the decoder to adapt to environmental changes. Thus, our framework can fine-tune and keep our RF tomography DNN model up-to-date for underground tuber sensing in a dynamic environment, where the multipath effects keep changing. The overall architecture of our framework is shown in Fig.~\ref{fig:architecture}.

In addition, we design a visual-RF cross-modal sensing system called VR-Spin to help annotate the underground RF sensing dataset automatically. As shown in Fig.~\ref{fig:testbedlayout}, we first align potato tubers underground with corresponding tuber masks on the soil surface in our ``nesting containers''. Then we use a camera to capture images of tuber masks and feed images to an image segmentation model, i.e., the Segment Anything Model~\cite{kirillov2023segment} to provide the ground-truth of the locations and shapes of the tubers. To increase the diversity of the dataset, we put the nesting containers on a platform and rotate the platform with various angles, so that root tubers can be located at various positions and orientations while recording the camera and RF network data. The idea of using a rotating cross-modality sensing system is inspired by data augmentation~\cite{dunlap2024diversify} commonly used in generating large amounts of data after the data acquisition stage. Our VR-Spin system enables data augmentation for RF sensing during the data acquisition stage. 

Finally, we use our VR-Spin system to collect RGB images and RSS measurements from 26 potato tubers in different dynamic environments. We build an underground potato sensing dataset to evaluate our DRIFT framework. Extensive experimental results show that our approach achieves an average equivalent diameter error of 2.29 cm, over 20\% more accurate than the existing alternative approach. To summarize, this paper makes the following contributions. 
\vspace{-0.4em}
\begin{itemize}[leftmargin=*]
\item We propose a \textbf{d}ata-driven \textbf{r}ad\textbf{i}o \textbf{f}requency \textbf{t}omography (DRIFT) framework for detecting underground targets, e.g., root tubers, in dynamic environments. We design a statistical model to detect environmental changes, and propose a DNN model that includes a one-shot fine-tuning strategy to adapt to a new environment.
\vspace{-0.4em}
\item We design a cross-modal sensing system, in which a camera and an image segmentation model are used to automatically annotate the ground-truth of underground targets. We perform measurement campaigns with our system to build an underground tuber sensing dataset. 
\vspace{-0.4em}
\item Extensive experimental results show that DRIFT achieves an average of 23.2\% improvement over the SOTA approach at various dynamic environments.
\end{itemize}
\vspace{-0.4em}
\section{Background and Related Work}
We now introduce RF tomography and related work.

\subsection{RF tomography background}

We assume the sensing area of an RF tomography network is represented by an image vector $\mathbf{r}=[r_0,\cdots,r_{N-1}]^T$, where $N$ is the pixel number of the image vector, and $r_m$ is a measure of the current presence of our target, i.e., root tuber, in pixel $m$. Then, RF link measurements can be formulated as:
\begin{equation}\label{eq:forward}
    \mathcal{\mathbf{g}} = \mathbf{\mathcal{H}}(\mathbf{r}) + \mathbf{n},
\end{equation}
where $\mathcal{\mathbf{g}} = [g_0, \cdots g_{M-1}]^T $ is a vector with the dimension of $M$ that represents RF signals, e.g., RSS measurements. $\mathbf{n} \in \mathbb{R}^M$ is a noise term, and $\mathbf{\mathcal{H}}$ denotes the observation function.
The goal of RF tomography is to estimate $\mathbf{r}$ from the network link measurements $\mathbf{g}$. 
If the observation function can be approximated by a linear function, the root tuber image vector $\mathbf{r}$ can be estimated by solving an inverse problem~\cite{WilsonP10,zhao2014robust}. In this paper, we follow the data-driven approach and propose to train a DNN model $\mathcal{F}: \mathbf{g} \rightarrow \mathbf{r}$ to estimate the image vector $\mathbf{r}$: 
\begin{equation}
    \hat{\mathbf{r}} = \mathcal{F}(\mathbf{g};\Theta),
\end{equation}
where $\hat{\mathbf{r}}$ represents the reconstructed target cross-section image, and $\Theta$ represents the set of DNN model parameters.

\subsection{Related Work}
\emph{RF sensing.}
Various RF sensing techniques have been proposed for smart agriculture, food and forestry applications. 
For example, \cite{lu20223d} uses an GPR to detect underground plant roots. Although GPR-based methods demonstrate satisfactory performance in below-ground root sensing, their broader applicability is hindered by cost constraints, high power requirements, and substantial physical dimensions of the GPR. 
More recently, low-cost wireless devices have been proposed for RF sensing. 
For example, WiFi channel state information (CSI) and Sub-Terahertz signals are used for fruit ripeness sensing~\cite{liu2021wi, agritera}. 
RF tomography and machine learning algorithms are used to detect potato root tubers~\cite{wang2024demo} and estimate the moisture level of rice~\cite{almaleeh2022inline}.

\emph{Continual learning.}
Unlike conventional machine learning models built on static data distributions, continual learning is characterized by learning from dynamic data distributions to deal with real-world dynamics. A major challenge in continual learning is the catastrophic forgetting problem, in which adaptation to new distributions results in a reduced ability to capture old ones~\cite{wang2024comprehensive}. 
In RF sensing, an effective way is to fine-tune a pre-trained model using limited samples. For example, to mitigate the challenges faced by millimeter-wave radio-based gesture recognition in heterogeneous environments, \cite{liu2022mtranssee} designs an innovative approach, which allows for practical gesture recognition using pre-learned experiences with limited target samples for fine-tuning. 
However, we have found little work applying various continual learning methods to RF tomography sensing, especially for underground target detection. 

\section{Methods and Systems} \label{S:algorithm-model}
We describe the cross-modal sensing system VR-Spin and the DRIFT framework in this section. 

\subsection{Overview}
The overview of the DRIFT framework is shown in Fig.~\ref{fig:architecture}.
First, RSS measurements from an RF tomography network are fed into the data imputation module for processing missing data due to RF interference. Then the pre-processed data are fed into our statistical model-based environment change detector, which uses signals from a sliding window to detect variations. If no environmental change is detected, the DNN model with the attention and convolution (ANC) modules will be used in model training, as shown in the top branch of Fig.~\ref{fig:architecture}. Once environmental change is detected, our DNN model is combined with a one-shot fine-tuning strategy for image reconstruction, as shown in the bottom branch of Fig.~\ref{fig:architecture}. 
Finally, the DRIFT framework also includes a post-processing module that applies image post-processing modules to filter out the remaining noise in the reconstructed images.

\subsection{Environment Change Detection}
For environmental change detection, we observe that under normal conditions, RSS data from most of our wireless links remain stable. However, dynamic events can cause significant variations in RSS values. Thus, we first calculate RSS statistics under normal and dynamic conditions. Specifically, we use a sliding window on the RSS time series data, and at a particular time, we calculate the standard deviation for each link within the moving window. Once these standard deviations are obtained, we rank them in descending order. Then we calculate the average of the standard deviations $\bar{\sigma}$ from the top-k links. For RSS data collected from a static environment without much motion interference, we follow the procedures above to calculate the standard deviation parameter $\bar{\sigma}_{static}$ at a static environment and use it as the noise level metric. During experiments in a dynamic environment, we use parameter $\bar{\sigma}_{static}$ as the threshold to detect environment change. 

\subsection{DNN Model with One-shot Fine-tuning} \label{sec:anc}
For our RF tomography DNN model, we propose a decoder comprising attention and convolution layers. The attention layer can adaptively adjust features, enhancing target-related features while attenuating those unrelated to the target, i.e., root tuber. Specifically, the feature from the multi-branch CNN is flattened and fed into ANC for imaging. After a linear layer, the output is reshaped into a two-dimensional feature map, which is refined by an attention layer implemented as a weight matrix. The feature map is then upsampled via bilinear interpolation and smoothed using convolutional layers. After training the neural network, we use the canny algorithm~\cite{canny} as a post-processing module to detect the edges of the underground tuber and treat the surrounding area as the tuber region.

Furthermore, short-term dynamic changes caused by human activity or environmental variation lead to noticeable variations in RSS values, while the size and shape of the underground tuber remain unchanged. This means that we can adjust the model parameters using the latest data without the need for re-labeling, enabling the model to adapt to new environments. Based on this observation, we propose to use a one-shot fine-tuning method to update the parameters of the pre-trained model and maintain robust performance. Specifically, once an environmental change is detected, the RSS data collected from a root tuber after the dynamic change are used to fine-tune the linear layer in ANC, while the remaining model parameters are fixed. The fine-tuning process is supervised, meaning that the loss between the prediction and the cross-section label image of the tuber is calculated to guide the update of the pre-trained model.

\subsection{Cross-modal sensing system} \label{S:spin}
Our visual-RF sensing system, VR-Spin, includes the following hardware and software components. 

\begin{figure}[!t]
	\centering
	\subfloat[Spin testbed]{
		\centering
        \includegraphics[width=0.45\linewidth]{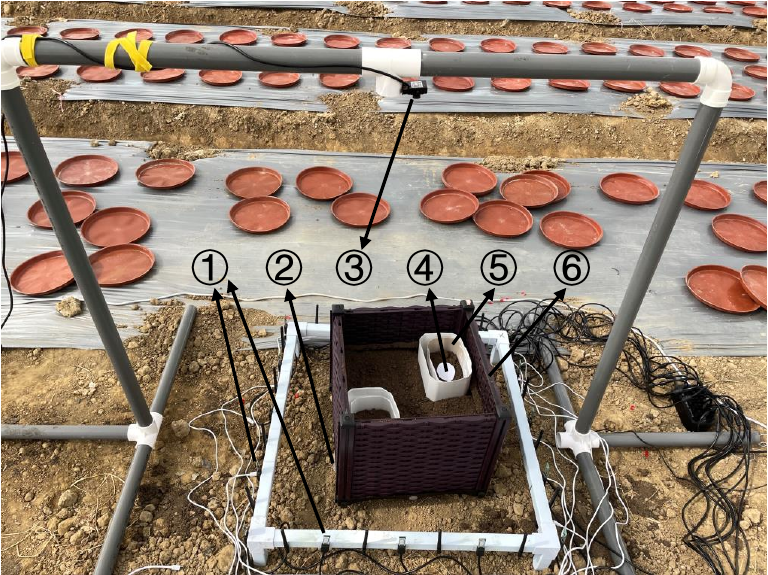}
        \label{fig:testbed}
	}
    \quad
	\centering
	\subfloat[Experimental Layout]{
		\centering
       \includegraphics[width=0.45\linewidth]{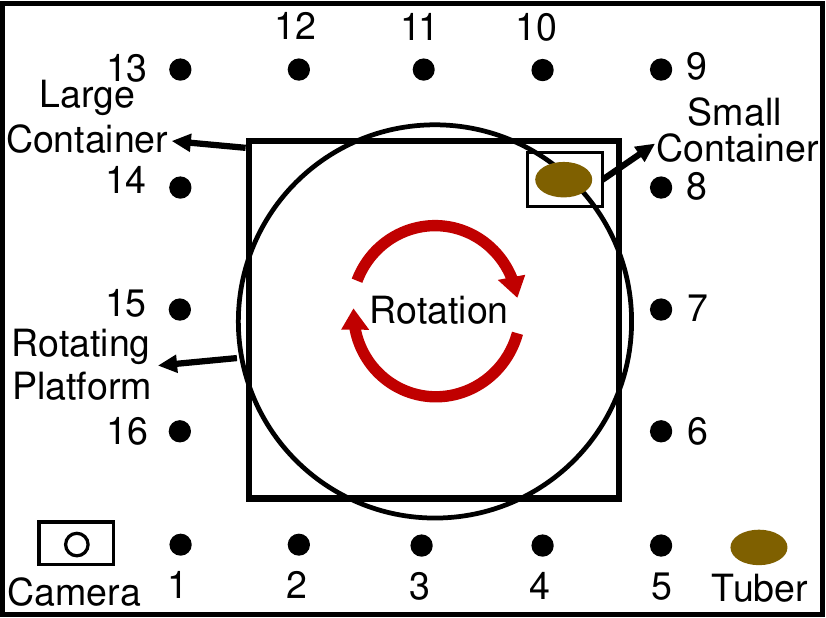}
        \label{fig:layout}
	}
	\caption{Our VR-Spin testbed includes an RF network with 16 nodes and a through-soil sensing toolkit featuring ``plug-and-play'' functionality and data augmentation capability. \ding{172} TI CC2531 nodes, \ding{173} Rotating platform, \ding{174} RGB camera, \ding{175} Marker with the same dimension as the tuber. \ding{176} Small container with the tuber. \ding{177} Larger container. 
 }
	\label{fig:testbedlayout}
\end{figure}

\emph{RF tomography network.} 
The RF tomography system was first developed in \cite{WilsonP10}. In this work, our RF tomography network is composed of 16 TI CC2531 nodes, as shown in Fig.~\ref{fig:testbedlayout}(a). Each wireless node is programmed with a multi-channel time-division multiple access communication protocol, and operates on one of the 16 frequency channels of the 2.4 GHz ISM band at a particular time. In our experiments, a designated sink node receives all packets transmitted by nodes and records them on a laptop. 

\emph{Through-soil sensing toolkit.}
Collecting large amounts of wireless sensing data is labor-intensive, especially for root tubers buried in soil. Thus, we design a set of tools including a rotating platform, and ``plug-and-play'' containers to facilitate data collection. First, we use two types of containers to avoid frequent soil digging and tuber burying. As shown in Fig.~\ref{fig:testbedlayout}(b), the large container contains soil with predefined locations to insert one or two small containers. Potato tubers with different dimensions are buried in different small containers with soil, allowing for easy replacement of tubers during data collection. Second, by placing the ``nesting containers'' on a rotating platform and rotating the platform with various predefined angles, root tubers can be located at various positions and orientations inside the sensing area. In addition, as shown in Fig.~\ref{fig:testbedlayout}(a), we place a marker with the same dimension as the underground tuber on the soil surface, to indicate the position and dimension of the tuber within the sensing area. We use a camera and an image segmentation model to capture the marker and automatically generate the ground truth for potato tubers.

\begin{table*}[!t]
    \centering
    \caption{Performance of DNN models in a dynamic environment. $E_1 \sim E_4$ denote different environmental conditions.}
    \begin{adjustbox}{width=0.85\linewidth}
    \begin{tabular}{cccccccccccccccc}
        \toprule
        \multirow{2}{*}{Test} & \multirow{2}{*}{Method} & \multicolumn{3}{c}{Leave-1-Out} & \multicolumn{3}{c}{Leave-2-Out} & \multicolumn{3}{c}{Leave-3-Out}  & \multicolumn{3}{c}{Leave-4-Out} \\
        \cmidrule(lr){3-5} \cmidrule(lr){6-8} \cmidrule(lr){9-11} \cmidrule(lr){12-14}
         &  & RPD  & IoU & EDE & RPD & IoU& EDE & RPD & IoU &EDE & RPD & IoU & EDE\\
        \midrule
        \multirow{3}{*}{$E_1 \rightarrow E_1$} & MC-LIM-UNet~\cite{wang2024demo} & 0.15 & 0.85& 2.96 & 0.14& 0.89 & 2.57 & 0.17 & 0.86 & 2.76 & 0.17 & 0.85& 2.82\\
         
         & CNN-LSTM~\cite{AttenConv} & 0.17 & 0.81 & 3.15 & 0.21 & 0.82 & 3.18 & 0.22 & 0.82 & 3.21 & 0.26 & 0.80 & 3.48 \\

         & \textbf{Ours} & \textbf{0.14} & \textbf{0.84} &\textbf{2.87}& \textbf{0.10} & \textbf{0.90} & \textbf{2.22} &  \textbf{0.14} &  \textbf{0.87} & \textbf{2.39}&  \textbf{0.20} &  \textbf{0.86} &\textbf{2.81} \\
        \midrule
        \multirow{3}{*}{$E_1 \rightarrow E_2$} & MC-LIM-UNet~\cite{wang2024demo}& 0.11 & 0.90 & 2.53 & 0.40 & 0.73 & 4.17 & 0.50 & 0.68 & 4.38 & 0.43 & 0.72 & 4.16\\
         
         & CNN-LSTM~\cite{AttenConv} & 0.10 & 0.88 & 2.46 & 0.24 & 0.81& 3.29 & 0.24 & 0.80 & 3.35 & 0.39 & 0.74 & 3.94 \\
         
         & \textbf{Ours} & \textbf{0.15} & \textbf{0.83} & \textbf{2.94} &  \textbf{0.08} & \textbf{0.91} &\textbf{1.98} &  \textbf{0.14} &  \textbf{0.88} &\textbf{2.38}& \textbf{0.13} &  \textbf{0.88} &\textbf{2.37}\\
        \midrule
        \multirow{3}{*}{$E_2 \rightarrow E_3$} & MC-LIM-UNet~\cite{wang2024demo}& 0.16 & 0.84 &3.06 & 0.28 & 0.76 &3.41 & 0.32 & 0.75 & 3.54 & 0.29 & 0.77 &3.39\\
         
         & CNN-LSTM~\cite{AttenConv} & 0.13 & 0.86 &2.73 & 0.18 & 0.85& 3.07 & 0.14 & 0.88 &2.61 & 0.22 & 0.84 &3.15\\
         
         & \textbf{Ours} & \textbf{0.16} & \textbf{0.83} & \textbf{3.04} & \textbf{0.05} & \textbf{0.90} &\textbf{1.55} &  \textbf{0.13} &  \textbf{0.88} &\textbf{2.27} &\textbf{0.16} &  \textbf{0.88} &\textbf{2.48}\\
        \midrule
        \multirow{3}{*}{$E_3 \rightarrow E_4$} & MC-LIM-UNet~\cite{wang2024demo} & 0.15 & 0.84 & 2.99 & 0.28 & 0.74 &3.43 & 0.31 & 0.76  &3.49 & 0.26 & 0.83 &3.18\\
         
         & CNN-LSTM~\cite{AttenConv} & 0.12 & 0.87 &2.70 & 0.16 & 0.85  &2.87 & 0.26 & 0.80 &3.49 & 0.22 & 0.84 & 3.12\\

         & \textbf{Ours} &\textbf{0.15} &  \textbf{0.83} &\textbf{2.99} &  \textbf{0.06} & \textbf{0.90} & \textbf{1.66} &  \textbf{0.13} &  \textbf{0.88} &\textbf{2.25} & \textbf{0.10} &  \textbf{0.89} &\textbf{2.04}\\
        \bottomrule
    \end{tabular}
    \end{adjustbox}
    \label{tab:dyevaluation}
\end{table*}

\emph{RGB camera for ground truth.}
We follow the procedures below to automatically annotate the ground truth of the potato tuber cross-section using our VF-Spin system. First, the potato tuber is placed horizontally in a smaller container with soil, ensuring that its maximum cross-section remains parallel to the ground. One end of four sticks is inserted into the soil and fixed around the tuber, with the other end exposed above the soil surface, as shown in Fig.~\ref{fig:testbedlayout}(a). A marker with the same dimension as the potato tuber is placed on the soil surface within the region surrounded by the sticks, indicating the position and dimension of the underground potato tuber. When the potato tuber and container are rotated on the platform, a camera at a fixed location captures RGB images. An image segmentation algorithm~\cite{kirillov2023segment} is then used to segment the pixels of the marker in the RGB image. Second, we establish a two-dimensional coordinate system for the sensing area, and the pixels corresponding to the marker in the RGB image are converted into coordinates within this system, representing the coordinates of the tuber in the sensing area. Third, we construct a ground truth image, where each pixel corresponds to a region in the two-dimensional coordinate system. Pixels corresponding to the coordinates of the tuber are assigned a value of 1, while all other pixels are assigned a value of 0. Using the tools and methods described above, we collect a large dataset of potato tubers with varying dimensions and shapes from different locations and environment conditions.

\emph{Data imputation.} 
To address the missing data issue caused by wireless interference, we develop a data pre-processing module that performs data imputation using the nearest packets. For example, when a wireless node fails to receive packets from other nodes on a frequency channel, we impute the missing RSS values based on the most recent data on the same frequency channel. Similarly, if all packets on a particular channel are lost, the most recent values on the same channel will be used in data imputation. 

Finally, we make our wireless potato sensing (WPS) dataset publicly available at IEEE Dataport~\cite{wps-dataset}. The code of our DRIFT framework is also available on https://github.com/Data-driven-RTI/DRIFT.

\section{Experiments and Results}
\subsection{Dataset} \label{sec:collection}
Our RF tomography underground sensing dataset can be divided into four subsets. The first subset contains RSS data from 26 fixed-position potato tubers in a static environment. The second, third, and fourth datasets focus on environmental changes, commonly occurring in real-world scenarios.
In our experiments, the RF tomography network sensing area is set to 72 cm by 72 cm, and data is collected from 26 potato tubers, which have lengths ranging from 10.5 cm to 2 cm and widths ranging from 7 cm to 1 cm. First, RSS data is collected from 26 potato tubers placed in a predefined position within the initial environment to build the pre-trained model. These potato tubers are placed at depths ranging from 10.7 cm to 15.5 cm, corresponding to their different sizes. Then, we randomly select 5 root tubers of varying sizes to collect RSS measurements in dynamic changes, enabling the fine-tuning and evaluation of our RF tomography model. During data collection, we continuously change the layout of the environment by moving furniture around the sensing area. Concurrently, human activities, such as walking and replacing small containers, will also create multipath effect changes. We use $E_1 \sim E_4$ to represent 4 continuously changing environments, respectively: $E_1$ represents the initial environment condition, $E_2$ represents the environment after moving a paper box, $E_3$ represents the environment after placing a chair around the sensing area, and $E_4$ represents the environment after switching the positions of the chair and the paper box. 
Note that we choose $K$ number of potato tubers in our leave-k-out evaluation, in which a total number of $26-K$ tubers are used in the initial model training and one tuber is used in the model fine-tuning. 

\subsection{Metrics and Baselines}

\emph{Evaluation metrics.}
We use the following metrics to evaluate our approach: intersection over union~(IoU), equivalent diameter error~(EDE) and relative pixel difference (RPD). IoU is used to quantify the similarity between an image and its ground truth. RPD measures the ratio of the difference in pixel count between the estimated and ground truth to the total pixel count of the ground truth. EDE is calculated as the diameter of a circle whose area is equal to the difference between the estimate and the ground truth. 

\emph{Baselines.} We choose two data-driven RF tomography methods as our baseline models~\cite{AttenConv,wang2024demo}.
First, \emph{CNN-LSTM~\cite{AttenConv}} develops an attention-based bidirectional convolutional LSTM model to perform image reconstruction. Second, \emph{MC-LIM-UNet~\cite{wang2024demo}} applies a neural network that integrates a multi-branch CNN with a UNet model for underground potato tuber image reconstruction. 
Note that the inference of all the DNN models can be performed in real time on regular computing platforms. 

\subsection{Evaluation on Dynamic Environments}
As mentioned in Section~\ref{sec:collection}, we create dynamic environments ($E_2$ - $E_4$) on purpose by walking inside the RF sensing area and changing the environmental layout to evaluate the robustness of our method. 
In our evaluation, we use RSS data from $E_1$ to build an initial model, which is fine-tuned and evaluated on the rest of the experiments $E_2 \sim E_4$. 
Specifically, we use one potato tuber to fine-tune our pre-trained model, and use the remaining $K \in {1,2,3,4}$ number of potato tubers in our leave-k-out test to evaluate the performance of our DRIFT framework. 

\begin{figure}[!t]
	\centering
	\subfloat[from MC-LIM-UNet]{
		\centering
        \includegraphics[width=0.45\linewidth]{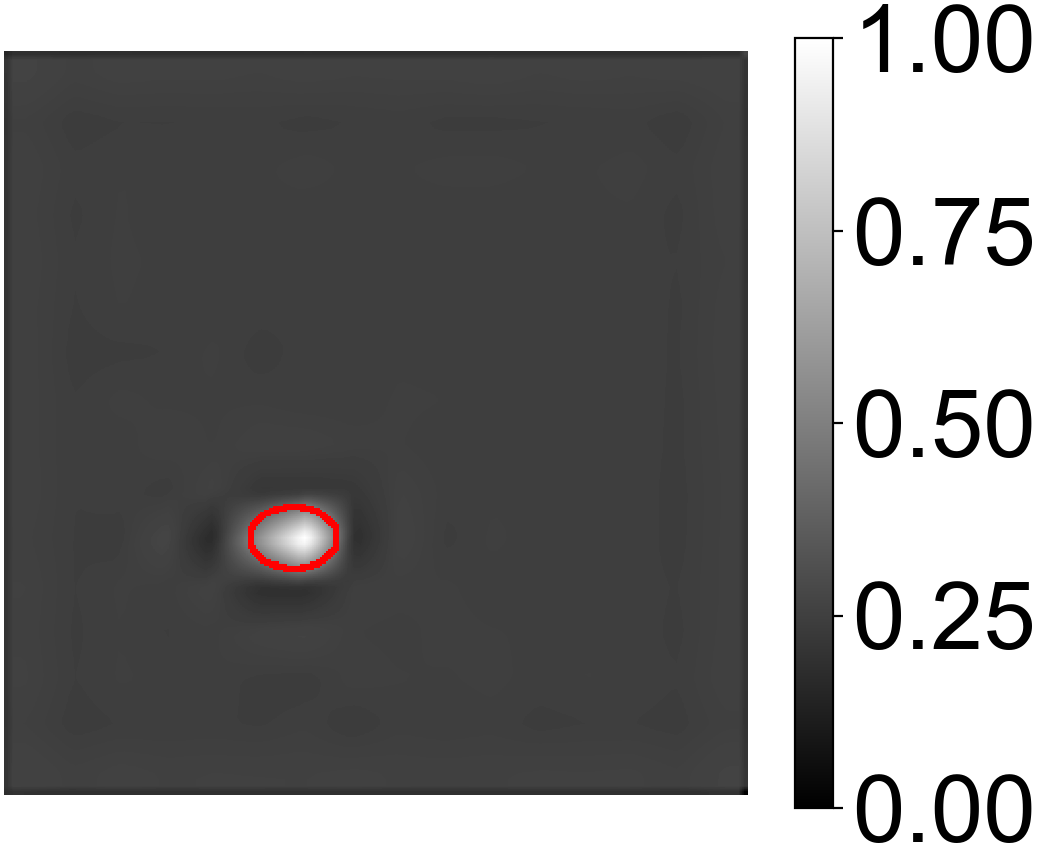}
        \label{fig:RPD}
	}
    \quad
	\centering
	\subfloat[from our model]{
		\centering
       \includegraphics[width=0.45\linewidth]{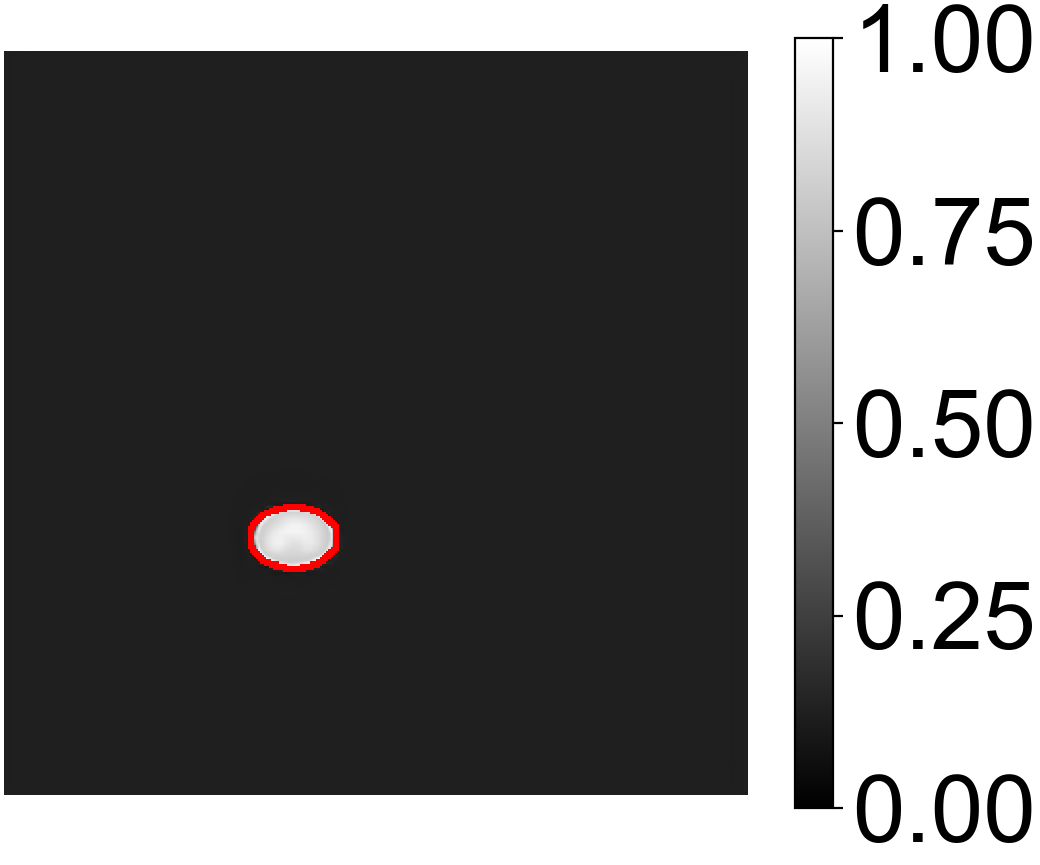}
        \label{fig:iou}
	}
	\caption{Imaging results at a dynamic environment.}
	\label{fig:results}
\end{figure}

Table~\ref{tab:dyevaluation} shows the RPD, IoU and EDE values of our model in comparison with baselines. 
From Table~\ref{tab:dyevaluation}, we see that for the initial static environment $E_1$, our model does not significantly outperform baselines. 
In addition, since only one potato tuber is used in the leave-1-out, certain evaluation metrics from our model are even lower than our baselines. 
For example, for the leave-1-out test on $E_1$, the IoU metric from our model is 0.84, similar to the MC-LIM-UNet model. 
However, as the number of tubers $K$ increases in leave-2-out, leave-3-out and leave-4-out tests, our model outperforms the baselines in all cases.  
Our method achieves average RPD and IoU values of 0.07 and 0.90 at the leave-2-out experimental setting, respectively. These values outperform those reported by MC-LIM-UNet~\cite{wang2024demo}~(0.28 and 0.78) and CNN-LSTM~\cite{AttenConv}~(0.17 and 0.83). Additionally, our method achieves an average EDE value of 1.85 cm at the leave-2-out experimental setting, compared to 3.40 cm and 2.95 cm reported by MC-LIM-UNet and CNN-LSTM, respectively. On average, our approach achieves an average equivalent diameter error of 2.29 cm, 23.2\% more accurate than the MC-LIM-UNet model. Finally, we show potato tuber cross-section image reconstruction results in Fig.~\ref{fig:results}. We see that the MC-LIM-UNet model produces a blurry image while our model provides more accurate cross-section area compared with the ground-truth indicated by the red circles. Note that we also investigate the impact of the number of nodes in an RF tomography network, with results included in our supplementary material. 

\section{Conclusion}
In this work, we propose a data-driven RF tomography framework, which includes an environment change detector and an DNN model with one-shot fine-tuning capability to achieve robust tomographic imaging of underground tubers at dynamic environments. We design a cross-modal visual and RF sensing system to automatically annotate the ground-truth of underground targets. 
We further build an underground potato tuber RF sensing dataset with our VR-Spin system. Extensive evaluation shows that our approach achieves over 23\% improvement for various dynamic environments. 
We believe that this work can pave the way for future study of applying other continual learning techniques to further improve data-driven underground RF sensing. 

{\footnotesize
\bibliographystyle{ieee}
\bibliography{sample-base}

\begin{thebibliography}{10}\itemsep=-1pt

\bibitem{wps-dataset}
{IEEE Dataport: Underground Root Tuber Sensing with Wireless Networks}, 2025.
\newblock (Accessed May-25-2025).

\bibitem{agritera}
S.~S. Afzal, A.~Kludze, S.~Karmakar, R.~Chandra, and Y.~Ghasempour.
\newblock Agritera: Accurate non-invasive fruit ripeness sensing via
  sub-terahertz wireless signals.
\newblock In {\em Proceedings of the 29th Annual International Conference on
  Mobile Computing and Networking}, pages 1--15, 2023.

\bibitem{almaleeh2022inline}
A.~A. Almaleeh, A.~Zakaria, L.~M. Kamarudin, M.~H.~F. Rahiman, D.~L. Ndzi, and
  I.~Ismail.
\newblock Inline 3d volumetric measurement of moisture content in rice using
  regression-based ml of rf tomographic imaging.
\newblock {\em Sensors}, 22(1):405, 2022.

\bibitem{canny}
J.~Canny.
\newblock A computational approach to edge detection.
\newblock {\em IEEE Transactions on pattern analysis and machine intelligence},
  (6):679--698, 1986.

\bibitem{dunlap2024diversify}
L.~Dunlap, A.~Umino, H.~Zhang, J.~Yang, J.~E. Gonzalez, and T.~Darrell.
\newblock Diversify your vision datasets with automatic diffusion-based
  augmentation.
\newblock {\em Advances in Neural Information Processing Systems}, 36, 2024.

\bibitem{fu2020environment}
X.~Fu, G.~Fortino, P.~Pace, G.~Aloi, and W.~Li.
\newblock Environment-fusion multipath routing protocol for wireless sensor
  networks.
\newblock {\em Information Fusion}, 53:4--19, 2020.

\bibitem{gano2024drone}
B.~Gano, S.~Bhadra, J.~M. Vilbig, N.~Ahmed, V.~Sagan, and N.~Shakoor.
\newblock Drone-based imaging sensors, techniques, and applications in plant
  phenotyping for crop breeding: A comprehensive review.
\newblock {\em The Plant Phenome Journal}, 7(1):e20100, 2024.

\bibitem{kirillov2023segment}
A.~Kirillov, E.~Mintun, N.~Ravi, H.~Mao, C.~Rolland, L.~Gustafson, T.~Xiao,
  S.~Whitehead, A.~C. Berg, W.-Y. Lo, et~al.
\newblock Segment anything.
\newblock In {\em Proceedings of the IEEE/CVF international conference on
  computer vision}, pages 4015--4026, 2023.

\bibitem{liu2022mtranssee}
H.~Liu, K.~Cui, K.~Hu, Y.~Wang, A.~Zhou, L.~Liu, and H.~Ma.
\newblock Mtranssee: Enabling environment-independent mmwave sensing based
  gesture recognition via transfer learning.
\newblock {\em Proceedings of the ACM on Interactive, Mobile, Wearable and
  Ubiquitous Technologies}, 6(1):1--28, 2022.

\bibitem{liu2021wi}
Y.~Liu, L.~Jiang, L.~Kong, Q.~Xiang, X.~Liu, and G.~Chen.
\newblock Wi-fruit: See through fruits with smart devices.
\newblock {\em Proceedings of the ACM on Interactive, Mobile, Wearable and
  Ubiquitous Technologies}, 5(4):1--29, 2021.

\bibitem{lu20223d}
Y.~Lu and G.~Lu.
\newblock 3d modeling beneath ground: Plant root detection and reconstruction
  based on ground-penetrating radar.
\newblock In {\em Proceedings of the IEEE/CVF Winter Conference on Applications
  of Computer Vision}, pages 68--77, 2022.

\bibitem{roitsch2019new}
T.~Roitsch, L.~Cabrera-Bosquet, A.~Fournier, K.~Ghamkhar, J.~Jim{\'e}nez-Berni,
  F.~Pinto, and E.~S. Ober.
\newblock New sensors and data-driven approaches—a path to next generation
  phenomics.
\newblock {\em Plant Science}, 282:2--10, 2019.

\bibitem{wang2024comprehensive}
L.~Wang, X.~Zhang, H.~Su, and J.~Zhu.
\newblock A comprehensive survey of continual learning: Theory, method and
  application.
\newblock {\em IEEE Transactions on Pattern Analysis and Machine Intelligence},
  2024.

\bibitem{wang2024demo}
T.~Wang, Y.~Zhao, J.~Liu, and Y.~Zhuang.
\newblock Underground potato root tuber sensing via a wireless network.
\newblock In {\em 2024 23rd ACM/IEEE International Conference on Information
  Processing in Sensor Networks (IPSN)}, pages 251--252. IEEE, 2024.

\bibitem{WilsonP10}
J.~Wilson and N.~Patwari.
\newblock Radio tomographic imaging with wireless networks.
\newblock {\em {IEEE} Trans. Mob. Comput.}, 9(5):621--632, 2010.

\bibitem{wilson2010see}
J.~Wilson and N.~Patwari.
\newblock See-through walls: Motion tracking using variance-based radio
  tomography networks.
\newblock {\em IEEE Transactions on Mobile Computing}, 10(5):612--621, 2010.

\bibitem{AttenConv}
H.~Wu, X.~Ma, C.~H. Yang, and S.~Liu.
\newblock Attention based bidirectional convolutional {LSTM} for
  high-resolution radio tomographic imaging.
\newblock {\em {IEEE} Trans. Circuits Syst. {II} Express Briefs},
  68(4):1482--1486, 2021.

\bibitem{yang2023one}
C.~Yang, Y.~Shen, Z.~Zhang, Y.~Xu, J.~Zhu, Z.~Wu, and B.~Zhou.
\newblock One-shot generative domain adaptation.
\newblock In {\em Proceedings of the IEEE/CVF International Conference on
  Computer Vision}, pages 7733--7742, 2023.

\bibitem{zhao2014robust}
Y.~Zhao and N.~Patwari.
\newblock Robust estimators for variance-based device-free localization and
  tracking.
\newblock {\em IEEE Transactions on Mobile Computing}, 14(10):2116--2129, 2014.

\end{thebibliography}
}

\end{document}